\documentstyle[twocolumn,aps]{revtex}
\begin{document} 
\title{
\small
To be presented at "Experimental Workshop on High Temperature Superconductors and Related Materials 
(Advanced Activities)" to be held from October 19$^{th}$ to November 6$^{th}$, 1998 in S. C. Bariloche, 
Argentina.\\ 
\begin{flushright}
TIFR/CM/98/302(I)
\end{flushright}
\large
SIMULTANEOUS OBSERVATION OF FISHTAIL EFFECT AND PEAK EFFECT IN 2H-NbSe$_2$\\
\vskip 0.25truecm
\normalsize
S. S. Banerjee$^{1,*}$, S. Ramakrishnan$^1$ and A. K. Grover$^1$
G. Ravikumar$^2$, P. K. Mishra$^2$ and V. C. Sahni$^2$
P. L. Gammel$^3$, D. J. Bishop$^3$ and E. Bucher$^3$
S. Bhattacharya$^4$
\vskip 0.1truecm
\noindent
{\it $^1$Tata Institute of Fundamental Research,
Mumbai-400005, India\\
$^2$TPPED, Bhabha Atomic Research Center,
Mumbai-400085, India\\
$^3$Lucent Technologies, Murray Hill, NJ 07940, USA\\
$^4$ NEC Research Institute, 4 Independence
Way, Princeton, New Jersey 08540\\}
\bigskip
\small \vskip 0pt
\noindent \hfill 
We report on the discovery of the simultaneous observation of Fishtail Effect (FE) and Peak Effect (PE) via 
the study of angular dependence of dc magnetization hysteresis loops in a clean crystal of 2H-NbSe$_2$. 
These result clarify and establish the occurrence of the reentrant characteristics in order to disorder 
transformation in an isothermal scan close to zero field superconducting transition temperature of 2H-
NbSe$_2$. The ubiquitous FE arises from coalescence of two anomalous variations in current density 
J$_c$, one of which is related to the collapse of the rigidity FLL close to H$_{c2}$ and another one located 
at very low fields presumably corresponds to the pinning induced order to disorder transformation. When 
the two effects get well separated, an ordered state of FLL exists in the intermediate field region.
\small
\begin{flushleft}
PACS numbers :64.70 Dv, 74.60 Ge, 74.25 Dw, 74.60
Ec,74.60 Jg\\ 
* Corresponding author; e-mail: sb@tifr.res.in
\end{flushleft}
}
\maketitle
\normalsize
\noindent
The vortex state of type II superconductors conceived by Abrikosov \cite{r1} encompasses the 
translational symmetry and such an ideal flux line lattice (FLL) is expected to move with arbitrarily small 
transport current under the influence of Lorentz force. The chemical impurities and structural imperfections 
in the underlying atomic lattice in any real sample provide the sources for preferentially pinning the flux 
lines at the expense of destroying the perfect translational symmetry of the FLL and thereby impart a 
critical current density (J$_c$) to the vortex array. J$_c$ of any superconducting specimen is expected to 
monotonically decrease while approaching the superconducting to normal phase boundary as the field 
increases (at fixed T) or as the temperature increases (at fixed H). However, experimentally one encounters 
the phenomenon of anomalous maximum in J$_c$ \cite{r2,r3,r4,r5} with increase in H(or T) in all varieties 
of superconducting samples. This anomalous maximum in J$_c$ is referred to either as Peak Effect (PE) 
\cite{r6} or as Fishtail Effect (FE) \cite{r7}. The basis of the latter nomenclature namely, FE, is the {\it 
characteristic shape of the isothermal dc magnetization hysterisis loop} of a pinned type-II superconductor.

As the concept of pinning relates inversely to the notion of perfect periodicity in the underlying 
symmetry \cite{r8,r9} of FLL, it is now generally accepted that the anomalous maximum in J$_c$ 
elucidates the occurrence of order to disorder transformation \cite{r10,r11,r12} in the FLL as a consequence 
of competition and interplay between the elastic energy, pinning energy and thermal fluctuations \cite{r13}. 
The advent of high T$_c$ superconductors focused widespread attention \cite{r13} on the issue of thermal 
melting \cite{r14} of the pure Abrikosov FLL. In recent years, convincing data have been collated in favor 
of first order nature of melting of FLL in very pure and nearly stoichiometric single crystals of High T$_c$
cuprates \cite{r15,r16,r17,r18}. In the well studied YBa$_2$Cu$_3$O$_7$ system, there are evidences that 
a sharp PE, exists in juxtaposition to the FLL, melting 
transition \cite{r17} and this coincidence supports a widely held view that a sharp PE located at the edge of 
depinning transition of FLL is a finger print of collapse of the elastic moduli of the FLL. In slightly off-
stoichiometric single crystals of YBa$_2$Cu$_3$O$_7$ \cite{r3}, the maximum in J$_c$ has been reported 
to evolve continuously 
from a narrow peak lying close to H$_{c2}$ (i.e., the usual Peak Effect) to a broad hump extending over a 
large field region located far away from H$_{c2}$ (i.e., the Fishtail Effect) as the effective pinning increases 
as a consequence of either the increase in quenched random disorder or the decrease in temperature. Thus, 
notionally, the PE and the FE may be treated as {\it distinct} and {\it mutually exclusive} features in dc 
magnetization 
hysterisis data though intimately related. In the context of single crystals of Bismuth Cuprate system, in 
which 
there have been reports of only the fishtail type of anomalous maximum in J$_c$ \cite{r18}, it was 
demonstrated from $\mu$SR studies \cite{r19} that the spatial order of FLL undergoes a sharp change at a 
field value coincident with the onset of fishtail anomaly in dc magnetization hysteresis data. Another 
$\mu$SR \cite{r20} study in the same Bismuth Cuprate system in different (H, T) region had established 
that the spatial order of 
the FLL undergoes a sudden reduction at (H, T) values where melting of FLL was anticipated. Thus, both 
peak effect and fishtail effect could be associated with the notion of occurrence of order to disorder 
transformation which is either first order or second order. There have been numerous reports in high T$_c$ 
superconductors wherein different phase boundaries across which vortex state undergoes order to disorder 
transformations corresponding to onset and peak position of fishtail 
effect have been drawn in the (H, T) space.  Some of these phase boundaries match with those phase 
boundaries drawn from other 
transport and thermodynamic measurements whereas some others appear to approach them at multicritical 
points (see, for instance, Fig.4 in \cite{r3} and Fig.4 in \cite{r18}).

Considering that different phases of vortex matter and transformation amongst them are a consequence of 
competition and interplay between elastic energy , pinning energy and thermal fluctuations, it is of interest 
to explore the presence of such characteristics in the vortex states of conventional low temperature 
superconductors in the appropriate (H,T) region. Amongst the low temperature superconductors, the clean 
single crystals of 2H-NbSe$_2$ system having T$_c$(0) $\approx$ 7 K are in current focus \cite{r21}, from 
the point of view of pristine physics issues of vortex matter \cite{r13} because of (i) their very weak pinning 
nature, (ii) their relatively large Ginzburg number value (G$_i$ $\sim$ 10$^{-4}$) \cite{r4} and (iii) the 
existence of a robust PE in them \cite{r4,r22,r23}. We present here new results pertaining to the discovery 
of simultaneous presence of fishtail effect and peak effect in a clean crystal of 2H-NbSe$_2$ at low fields 
(in H $<$ 1 kOe where the FLL lattice constant a$_0$ $>$ 2000 A$^o$) and in the high temperature 
(T$/$T$_c$(0) $>$ 0.95) region. We 
believe that these results are the first of their kind. They pertain to a vortex array in its dilute limit where the 
interaction amongst the vortices is in a nascent stage and the fluctuation effects are very strong due to 
the close proximity to T$_c$(0). These results have lead us to construct phase boundaries in (H,T) space 
which have the potential to clarify complex issues emerging from studies related to PE/FE/FLL melting in 
high T$_c$ cuprates \cite{r13,r14,r15,r16,r17,r18}. One specific issue that the simultaneous observation of 
FE and PE aims to resolve is 
the {\it reentrant} characteristic of order-disorder phase boundary of FLL \cite{r5,r13,r14,r24} as the 
interaction gets progressively 
enhanced (with increasing field) while thermal fluctuations (T) and quenched random disorder remain fixed.

For our present study we chose a single crystal of 2H-NbSe$_2$ (T$_c$(0)$\sim$7.2K with a 
$\Delta$T$_c$$\sim$50 mK). The sample dimensions are 5 X 2 X 1 mm$^3$ and J$_c$ values in it lie 
below 
1000 A/cm$^2$. It is our belief that the present sample has a level of purity which lies in between those of 
the crystals A and B, used by us in our earlier studies \cite{r5}. We measured its ac magnetization 
screening response using a high sensitivity ac-susceptometer \cite{r25} and dc- magnetization using a 
standard 
Quantum Design SQUID magnetometer with a specially designed home made sample holder which allows 
an angular variation of  0$^o$ to 180$^o$ between the field H and the ab-plane of the single crystal 
 
Fig.1 shows the temperature dependence of the diamagnetic screening response ($\chi$$^{\prime}$(T)) for 
the FLL created at different H$_{dc}$ ($\parallel$ c) values. From a generalized Critical State Model 
\cite{r26} relation \cite{r27} 
:$$ \chi^{\prime}~ \approx~-1~+ {\alpha.h_{ac} \over J_c},~~~~~~~~~(1) $$, where $\alpha$ is a geometry 
and size 
dependent factor and J$_c$ is the (H,T)- dependent critical current density. From this  relationship, it 
can be deduced that PE phenomenon (i.e., peak in J$_c$) should manifest itself as an anomalous increase 
in the diamagnetic screening response ($\chi$$^{\prime}$(T)) for a given H$_{dc}$. In Fig. 1, we have 
marked with arrows the peak temperatures T$_p$ of the PE in $\chi$$^{\prime}$(T) behavior at various 
H$_{dc}$. It can be 
seen that T$_p$ values increase with decreasing H$_{dc}$ down to 200 Oe. Further more, another 
noteworthy feature in 
$\chi$$^{\prime}$(T) data is that, the PE starts 
to broaden substantially about T$_p$ value as one moves to lower fields (from 100 Oe to 50 Oe). To reveal 
the possible connection between the phenomenon of broadening of PE and the behavior of 
FLL at low fields it may be pertinent to point out here that, in the field range from 100 Oe to 50 Oe where 
the PE starts to broaden, the values of 
a$_0$ vary from 4800 A$^o$ to about 6800 A$^o$. Recalling that $\lambda$$_c$~$\sim$~4550~A$^o$, in 
2H-NbSe$_2$ in the (H,T) region under consideration \cite{r29}, the vortex array is in the dilute limit 
(a$_0$ $>$ $\lambda$) for H $<$ 100 Oe and the flux lines are only weakly interacting , with the result the 
FLL is in a state which is easily susceptible to thermal fluctuations and pinning effects. The inset in Fig.1 
shows the PE curve which has been determined by picking out the field - temperature values at which the 
peak 
in PE occurs in $\chi$$^{\prime}$(T) data. The noteworthy features of PE are : (i) at high fields (H $>$ 200 
Oe) the 
PE curve tracks the H$_{c2}$(T) curve, such that both of them have a slope of $\approx$ 5 kOe$/$K, 
however at lower 
fields (H $<$ 100 Oe) the PE curve bends away from the H$_{c2}$(T) curve, (ii) the progressive increase in 
the size of the error bars 
on the data points indicates the commencement of process of broadening out of PE phenomenon, while 
approaching T$_c$(0). Our previous studies \cite{r5} had 
revealed that increased effective pinning broadened out the PE, we thus surmise that an interplay between 
pinning and thermal 
fluctuation effects at low H - high T determines the broadening feature in PE in Fig.1. This prompts to 
investigate the 
features of PE and its broadening through isothermal magnetization M(H) hysterisis loop measurements, 
where we study the progression of flux line lattice of varying a$_0$ under the influence of fixed thermal 
energy. To 
perform these isothermal M(H) measurements, we chose particularly those temperatures (close to 
T$_c$(0)), where the PE broadened and the PE curve seems to bend away from H$_{c2}$(T) (see Fig.1).

A superconductor shields itself from any change in external magnetic field by setting up currents equal to 
J$_c$(B), which in 
turn result in the magnetization M at a given field(H). When one obtains a dc magnetization hysterisis loop, 
the width of the magnetization hysterisis at a given field, i.e., $\Delta$M ($\Delta$M(H)$=$ 
M(H$\downarrow$)$-$M(H$\uparrow$)) may be taken as a measure of J$_c$. This implies that any non-
monotonic variation in the 
behavior of J$_c$(H) can, therefore, show up as an anomalous modulation in the width ($\Delta$M) of the 
magnetization 
hysterisis loop. The PE emerges thus as an 
anomalous bubble like anomaly superposed on the quasi reversible magnetization (M(H)) 
hysteresis loop (see PE  region centered around H$_p$ in the main panel of Fig.2(a)). In isothermal dc 
magnetization hysterisis data at T=6.95 K in 2H-NbSe$_2$, we observe that there is a well formed PE 
(anomalous M(H) hysterisis 
bubble) centered around H$_p$$\approx$1000 Oe (cf. Fig.2(a)) and this peak field value is consistent with 
the T$_p$(H) curve in Fig.1. In Fig.2(a) in the field 
range from 50 Oe to 250 Oe (see arrow marked H$_d$ in Fig.2(a)), there is an anomalous modulation in the 
width of the 
magnetization hysterisis loop. We can determine the J$_c$ using the relationship J$_c$ $\propto$ 
$\Delta$M$/$d, where d is the thickness of the sample. In the inset of Fig.2(a) the behavior of J$_c$ versus 
H at T=6.95 K is shown in dark square symbols on a semi-log plot. Such a J$_c$(H) plot shows an overall 
linear (i.e., J$_c$~$\propto$~exp($-$H)) behavior 
(examine the straight line passing through the data points). It can be easily viewed that in the field range of 
70 Oe $\leq$ H $\leq$ 250 
Oe, there is an anomalous behavior in J$_c$(H) which brings out the deviation from the overall linear 
behavior. We 
label the center of gravity of this anomaly in J$_c$ as H$_d$ at $\approx$ 100 Oe. The development of this 
anomaly at H$_d$ as a function of temperature is investigated by studying the M(H) hysterisis loop (cf. 
main panel of Fig.2(c)) at T = 7.0 K, where the hysteresis loop seems to be completely anomalous and 
irreversible in the entire field range from 0 Oe $\leq$ H $\leq$ H$_{c2}$. The visible differences that emerge 
by comparing the M(H) hysterisis loop at T = 7.0 K and T = 6.95 K are : (i) one cannot precisely locate the 
low field anomaly at H$_d$ in the M(H) loop at 7.0 K and (ii) one cannot also clearly identify from the 
M(H) hysteresis curve the upper PE which occurs close to H$_{c2}$ . In the inset of Fig.2(c), we have put 
together the behavior of the J$_c$(H) at 7.0 K and 6.95 K. The usual expectation is J$_c$(H, T = 7.0K ) $<$ 
J$_c$(H, T = 6.95 K), however such an inequality is satisfied only for 
H~$\leq$~H$_d$. At H~$>$~H$_d$, the J$_c$(H, T = 7.0K ) $>$ J$_c$(H, T = 6.95 K ), which is 
a manifestation of an anomalous behavior in J$_c$ and it signals the occurrence of peak effect like feature in 
J$_c$ at low fields which are far from H$_{c2}$. It thus seems that the anomalous modulation in J$_c$ 
at H$_d$ at 6.95 K in Fig.2(a) is a fingerprint of a low field PE phenomenon and it presumably survives 
at 7.0 K. 

We now present an analysis of the pinning force F$_p$($=$J$_c$$*$H), which brings out features not 
readily apparent from the 
M(H) hysteresis data or from the J$_c$(H) behavior. We first determine F$_p$ using the J$_c$(H) data 
extracted from the M(H) hysterisis loops, and we call it F$_p$(measured). Next, 
we construct a monotonically decreasing J$_c$(H) which is devoid of any anomalous modulation, i.e., this 
constructed J$_c$ gets rid of any PE like features at low fields or at high fields (a procedure to construct 
such a 
J$_c$(H) can be seen from the semi log plot of J$_c$(H) (in the inset of Fig.2(a)) at 6.95 K). Using this 
constructed J$_c$, we determine the pinning 
force which we call F$_p$(constructed)=J$_c$(constructed)$*$ H. Finally, we determine 
$\Delta$F$_p$=F$_p$(measured)~$-$~F$_p$(constructed). $\Delta$F$_p$ is a quantity which can convey 
signatures of anomalous \cite{r29} changes in the intrinsic pinning force density being experienced by the 
flux lines. In Figs. 2(b), 2(d), 2(f), 2(h), we have plotted $\Delta$F$_p$ as a function of H. At 6.95 K (cf. 
Fig.2(b)), one can clearly distinguish two well resolved PE features in $\Delta$F$_p$(H). One PE located at 
H$_d$$\sim$100 Oe, is the novel low field PE,  this peak in $\Delta$F$_p$ at H$_d$ coincides with the 
field at which the anomalous non monotonic low field behavior in J$_c$(H) is observed (cf. inset of 
Fig.2(a)). The other PE, is the high field PE at H$_p$, which 
occurs close to H$_{c2}$. At T=7.0 K (cf. Fig.2(d)), one can again distinguish two well resolved peak like  
features 
in $\Delta$F$_p$,  at H$_d$ and H$_p$. It should be noted that although the M(H) hysteresis loop at 7.0K 
was devoid of any distinguishing features of PE at H$_d$ and H$_p$, through the present analysis we can 
now discern features of PE at both H$_d$ and H$_p$ in the $\Delta$F$_p$(H) plot at 7.0 K. If one now 
compares the Fig.2(b) with Fig.2(d), it seems that though the PE at H$_d$ does not move as we change T 
from 6.95 K to 7.0 K, the H$_p$ has moved from $\sim$1000 Oe at 6.95 K to $\sim$750 Oe at 7.0 K, which 
is a rate of 5 kOe/K. This slope value of dH$_p$/dT, matches well with the slope of the PE curve in the inset 
of Fig.1, which was determined from temperature dependent ac susceptibility measurements. From the 
above analysis one can 
conjecture that due to a motion to lower fields (close to H$_d$), of the PE at H$_p$ at the rate of 5 kOe/K 
as T is increased, one should be able to see the emergence of a single peak from the coalescence of two 
well resolved PE peaks in $\Delta$F$_p$ (i.e., one at H$_d$ and another at H$_p$). We now show the 
M(H) hysteresis loop at subsequently higher temperatures. The M(H) hysteresis loop at T=7.05 K (cf. Fig. 
2(e)), 
appears even more anomalous than that at 7.0 K. It is a broad, irreversible loop which 
results in a single broad peak in $\Delta$F$_p$ (cf. Fig. 2(f)). {\bf The M(H) hysterisis loop at 7.05 K now 
resembles so called ``Fishtail Effect Anomaly".} This 
fishtail effect like anomaly in the M(H) hysteresis loop and in $\Delta$F$_p$(H) plot persists at T $\geq$ 
7.05 K. Figs. 2(g) and 2(h) show M(H) hysterisis loop and the 
F$_p$(H) at T = 7.1 K. It is to be noted that the width of the FE anomaly, (i.e., $\Delta$H$_d$ in Fig.2(b), 
Fig.2(d), Fig.2(f), Fig.2(h)) has an interesting temperature 
dependence, we shall discuss it at a later stage.  We shall now 
demonstrate through a sequence of  M(H) hysterisis loops obtained at various orientations with respect to 
the ab plane of the single crystal as to how the composite broad FE effect at 7.0 K (see Fig.2(c)) gets 
resolved into two anomalous maxima in J$_c$ which are centered around H$_d$ and H$_p$.

It had been argued by Pippard \cite{r30} and also experimentally seen that the PE peak H$_p$ always 
scales with 
H$_{c2}$. 2H-NbSe$_2$ is an anisotropic system to which anisotropic Ginzburg-Landau description 
applies \cite{r31}. The angular dependence of H$_{c2}$ is given by as 
H$_{c2}$$=$H$_{c2}$($\parallel$c,T)(Sin$^2$($\theta$)+$\epsilon$$^2$Cos$^2$($\theta$))$^{-1/2}$, 
where $\epsilon$=1$/$(Anisotropy of crystal) and $\theta$ the angle between H and the ab plane of the 
single crystal of NbSe$_2$. As $\theta$ changes from 90$^o$ to 0$^o$, H$_{c2}$ increases from the value 
of H$_{c2}$($\parallel$c,T) to that of H$_{c2}$($\parallel$ab,T), and concomitantly the value of 
H$_p$ also follow suit. In Fig.3, the M(H) hysterisis loops in 2H-NbSe$_2$ at 7.0 K for different $\theta$ 
The insets in each of the Figs. 7(a) to (f) show 
F$_p$ vs H$/$H$_{c2}$($\theta$). In the inset of Fig.3(a) we can see that the low field, maximum in F$_p$ 
at 
H$_d$ and the high field PE peak at H$_p$ lie in juxtaposition. When the angle $\theta$ changes from 
90$^o$ towards 0$^o$, 
the following features in Fig.3 are noteworthy: (i) In all the insets of Fig.3, the peak in F$_p$ at H$_p$ 
seems to occur at H$/$H$_{c2}$($\theta$)$\approx$0.7. (ii) The low field peak in 
F$_p$ at H$_d$ shifts to lower values of H$/$H$_{c2}$($\theta$) as $\theta$ increases (iii) The most 
striking feature emerging from the plots of Fig.3 is the separating out of the two anomalous modulations in 
J$_c$ corresponding to a distinct double peak structure in 
the F$_p$(H) plot. The main panels of Fig. 3 show that by varying $\theta$, the 
PE peak at H$_p$ moves to higher fields (since H$_p$$/$H$_{c2}$($\theta$)$\approx$0.7),  and as the 
H$_p$ value 
moves sufficiently far away, the novel low field anomalous modulation in J$_c$ at H$_d$ gets identified 
distinctly as an independent feature. If  we attempt to fit the observed H$_{c2}$($\theta$) behavior to the 
above stated Ginzburg-Landau relationship, we get a value of $\epsilon$ as $\sim$0.5, or an anisotropy of 
about 2 for 2H-NbSe$_2$, which is a satisfactory 
value for NbSe$_2$ for H $<$ 2 kOe \cite{r28}. Thus our anisotropy study of the M(H) hysterisis loop 
elucidates the fact that, FE could be composed of an anomalous modulation in J$_c$ centered at low field 
H$_d$ and the usual PE peak at H$_p$. 

We now summarize our discussion and propose a schematic for the low H and high T part of the vortex 
phase diagram. 
In the inset of Fig.1(a), we had presented a phase diagram which comprised the PE phase boundary T$_p$ 
(H) across which the FLL disorders into an amorphous phase. However, this phase 
diagram is incomplete as it does not give us any indication of changes in the vortex matter occurring at field 
values such as at H$_d$ and $\Delta$H$_d$, which we can observe from the isothermal M(H) hysteresis 
loops. 
Returning to Fig.2, we consider the behavior of the width $\Delta$H$_d$ of the low field PE centered at 
H$_d$. Shown in Fig.4, is a shaded region which is obtained by estimating the full width at half maximum 
of the peak in $\Delta$F$_p$ at H$_d$ in $\Delta$F$_p$(H) plots (cf. Fig.2) at different T's. This high 
pinning shaded 
region is bounded by two curves H$_d$$^{lower}$ and H$_d$$^{upper}$, where $\Delta$H$_d$ = 
$|$H$_d$$^{upper}$ - H$_d$$^{lower}$$|$. The 
width $\Delta$H$_d$ seems to increase with T and reaches a maximum at about T = 7.05 K and above 
this temperature (cf. Fig.2), the width $\Delta$H$_d$ appears to decrease again. At very low H (a$_0$ $>$ 
$\lambda$), where the rigidity of the FLL (viz. c$_{66}$ $\sim$ exp(-a$_0$/$\lambda$)) is exponentially 
small, the FLL 
is highly susceptible to thermal fluctuations and pinning effects which could disorder the FLL. For vortex 
states corresponding to the shaded region of Fig.4 (low H - high T), the 
weak rigidity permits thermal fluctuations and pinning effects to dominate as evidenced by the broad hump 
in the M-H curves and hence the FLL states in the shaded region belong to ``disordered phase'' of vortex 
matter which is presumably characterized by a small Larkin volume V$_c$. We now attempt to explore 
features in our temperature dependent ac susceptibility and dc magnetization data which can possibly be 
linked to the 
phase labelled as the ``disordered phase'' in the H-T plane .  As implied from the above analysis, 
$\Delta$H$_d$ identifies the field span over which J$_c$(H) varies anomalously. In temperature dependent 
ac susceptibility data, the broadening of PE peak commenced in isofield runs at H $<$ 1kOe. We believe 
that the said broadening is a consequence of encountering the shaded region in fixed H scan (see Fig.4). On 
the other hand, in fixed T scan in dc magnetisation hysteresis data, the ``lower'' PE and ``upper'' PE appear 
as distinct features if H$_d$$^{upper}$ line lies sufficiently far below the H$_{c2}$ line. In the field range 
between the H$_d$$^{upper}$ and the PE region, the vortex state do not show any anomalous variation in 
J$_c$. We, therefore, feel that such FLL states identify the well ordered vortex solid. However in the field 
region above the point A (marked in main panel of Fig.4 ), where the 
shaded region and the H$_p$ line have merged, if one records an isothermal M(H) hysteresis loop, then 
this would be the region where the two PE would have merged and would have the appearance of the 
Fishtail like effect. Above H$_p$ and 
below H$_{irr}$ curves \cite{r5}, the FLL is presumably in an amorphous state with non zero pinning. 
Such amorphous state 
maybe indistinguishable in its pinning behavior from the ``disordered phase'' in the shaded region. Hence, 
we propose that the vortex region between H$_p$ and 
H$_{irr}$ lines and the shaded region labeled as the ``disordered phase" essentially identify the same vortex 
phase. In the 
inset of Fig.4, we show the phase boundary across which the ordered solid disorders. This has been 
constructed 
by combining the H$_p$ curve, to the H$_d$$^{upper}$ curve (see main panel of Fig.4). 
This new phase boundary across, designated as H$_{ord}$ in the inset of Fig.4, has a reentrant nature, i.e., 
in fixed T scan,  the order-disorder boundary shall be encountered twice. Such a reentrant characteristic is 
reminiscent of reentrant nature of the FLL melting curve of a pure (pinning free) Abrikosov state, first 
proposed by D. R. Nelson \cite{r13}. Between this newly constructed reentrant H$_{ord}$(T) line and the 
H$_{irr}$(T) line, the vortex matter can be designated as the ``pinned liquid phase", following the 
nomenclature in vogue in the literature on vortex state studies \cite{r13}. 

In conclusion we may state that 2H-NbSe$_2$, seems to be an appropriate system, whose intrinsic 
parameters, make the phase boundaries, 
across which the various phases of vortex matter exists, become accessible to within experimental limits. 
Though the present set of results are the first preliminary steps in characterizing the different phases of 
vortex matter, the experimental and theoretical investigations in this field are wide open to enormous 
possibilities and opportunities.  

We would like to acknowledge Nandini Trivedi, Satya Majumder and Mahesh Chandran for fruitful 
discussions.

\begin{figure}
\caption{Temperature variation of ac susceptibility (f=211 Hz, h$_{ac}$=1 Oe(r.m.s.)) in a single crystal of 
2H-NbSe$_2$ for H$_{dc}$ $\parallel$ c. The peak temperature T$_p$ have been identified by arrow. Note 
that PE peak broadens as H$_{dc}$ progressively decreases for H$_{dc}$ $<$ 1 kOe. The inset shows the 
plot of T$_p$(H) and T$_c$(H) values in the low field region of the magnetic phase diagram.}
\label{Fig.1}
\end{figure}
\begin{figure}
\caption{Isothermal dc magnetisation loops, extracting J$_c$ values and computed pinning force values in 
2H-NbSe$_2$ for H$_{dc}$$\parallel$c at various temperatures. For explanation of various symbols used 
in this figure, see text. }
\label{Fig.2}
\end{figure}
\begin{figure}
\caption{Angular dependence of magnetic hysteresis data in a crystal of 2H-NbSe$_2$ at 7.0 K. The angle 
$\theta$ is measured w.r.t. basal plane of hexagonal atomic lattice of 2H-NbSe$_2$. For explanation of 
various symbols in this figure, see text. }
\end{figure}
\begin{figure}
\caption{Magnetic phase diagram corresponding to temperature dependent ac susceptibility (Fig.1) and 
field dependent dc magnetisation hysterisis (Fig.1) data. The inset shows, schematically the proposed 
magnetic phase diagram constructed the results of ac and dc magnetisation experiments. See text for 
explanation of different phases boundaries.}
\label{Fig.4}
\end{figure}

\begin{references}
\bibitem{r29} Even in the absence of anomalous change in J$_c$(H), F$_p$(measured) is expected to 	      
show a peak as a function of H as it is a product of J$_c$ and H. However, such a 	      peak does not 
signal the existance of peak in J$_c$(H).
\end{references}
\end{document}